# Nonlinear Dipole Inversion (NDI) enables Quantitative Susceptibility Mapping (QSM) without parameter tuning

Daniel Polak[1,2,3], Itthi Chatnuntawech[4], Jaeyeon Yoon[5], Siddharth Srinivasan Iyer[2,6], Jongho Lee[5], Peter Bachert[1,7], Elfar Adalsteinsson[6], Kawin Setsompop[2,8,9] and Berkin Bilgic[2,8,9]

[1] Department of Physics and Astronomy, Heidelberg University, Heidelberg, Germany
[2] Department of Radiology, A. A. Martinos Center for Biomedical Imaging, Massachusetts General Hospital, Charlestown, Massachusetts, USA
[3] Siemens Healthcare GmbH, Erlangen, Germany
[4] National Nanotechnology Center, Pathum Thani, Thailand,
[5] Laboratory for Imaging Science and Technology, Department of Electrical and Computer Engineering, Seoul National University, Seoul, South Korea
[6] Department of Electronical Engineering and Computer Science, Massachusetts Institute of Technology, Cambridge, Massachusetts, USA
[7] Medical Physics in Radiology, German Cancer Research Center (DKFZ), Heidelberg, Germany
[8] Harvard Medical School, Boston, Massachusetts, USA
[9] Harvard-MIT Health Sciences and Technology, Massachusetts Institute of Technology, Cambridge, Massachusetts, USA

**Abbreviations**:
- CNN: convolutional neural network
- MEDI: morphology enabled dipole inversion
- NDI: nonlinear dipole inversion
- QSM: quantitative susceptibility mapping
- SMV: spherical mean value
- TGV: total generalized variation
- TKD: truncated k-space division
- TV: total variation
- VaNDI: variational nonlinear dipole inversion
- VN: variational network
- w.r.t: with respect to

**Acknowledgements:** This work was supported in part by NIH research grants: R01EB020613, R01EB019437, P41EB015896, U01EB025162, the shared instrumentation grants: S10RR023401, S10RR019307, S10RR019254, S10RR023043, and NVIDIA GPU grants.

# Abstract

We propose Nonlinear Dipole Inversion (NDI) for high-quality Quantitative Susceptibility Mapping (QSM) without regularization tuning, while matching the image quality of state-of-the-art reconstruction techniques. In addition to avoiding over-smoothing that these techniques often suffer from, we also obviate the need for parameter selection. NDI is flexible enough to allow for reconstruction from an arbitrary number of head orientations, and outperforms COSMOS even when using as few as 1-direction data. This is made possible by a nonlinear forward-model that uses the magnitude as an effective prior, for which we derived a simple gradient descent update rule. We synergistically combine this physics-model with a Variational Network (VN) to leverage the power of deep learning in the VaNDI algorithm. This technique adopts the simple gradient descent rule from NDI and learns the network parameters during training, hence requires no additional parameter tuning. Further, we evaluate NDI at 7T using highly accelerated Wave-CAIPI acquisitions at 0.5 mm isotropic resolution and demonstrate high-quality QSM from as few as 2-direction data.

# Introduction

Quantitative Susceptibility Mapping (QSM) provides exquisite gray/white matter contrast [1] and enables accurate quantification of iron in the brain [2]. It is also utilized to differentiate dia- and paramagnetic sources of contrast [3], detect tissue changes related to neuro-degenerative diseases [4]–[6] and estimate vessel oxygenation [7], [8]. However, it entails a difficult image reconstruction pipeline with several pre-processing steps, which are briefly summarized at the beginning of this contribution. This work will focus on the dipole inversion step, which aims to estimate the desired magnetic susceptibility from the acquired gradient echo phase information. Several benchmark and state-of-the-art techniques are described, and finally a new method for robust dipole inversion is introduced which mitigates some of the drawbacks of previous approaches.

Typically, the first step of the QSM reconstruction is the coil combination [9]–[12] of multi-channel receive data. Hereby, phase offsets between the different coil images are estimated to prevent destructive interference between complex signals. The resulting coil-combined phase images are initially wrapped into an interval of [-π, π]. To recover the underlying phase distribution of interest, a spatial image unwrapping technique (path-following: [13]–[16]; Laplacian: [17]–[19]) is applied, which usually relies on the assumption that the phase signal varies slowly from voxel to voxel. After unwrapping, the phase images are dominated by background phase effects which are commonly one to two orders of magnitude larger than those caused by the desired tissue susceptibility [20]–[22]. Several filtering techniques have been proposed [23]–[27] to remove unwanted phase contributions such as those caused by inhomogeneities of the static main magnetic field, macroscopic currents (MRI shim coils), or magnetic susceptibility variations outside the ROI (e.g. air-tissue interfaces) [28], [29]. In a final step, the tissue phase $\phi$ needs to be de-convolved with a known dipole kernel $d(k) = 1/3 - k_z^2/k^2$ to obtain the desired susceptibility $\chi$. Since the dipole kernel is not invertible, this inverse problem $\phi(k) = d(k)\chi(k)$ cannot be solved using a simple division in k-space but requires more sophisticated reconstruction techniques.

Truncated k-space Division (TKD) [30] inverts the dipole kernel directly $\chi(k) = \tilde{d}(k)\phi(k)$ where small absolute values in $d(k)$ are replaced by a constant number.

$$\tilde{d}(k) = \begin{cases} d(k)^{-1} & \text{if } \|d(k)\| > \delta \\ \text{sgn}(d(k)) \cdot \delta^{-1} & \text{otherwise} \end{cases}$$

However, the modification of the dipole kernel may result in systematic underestimation of the tissue susceptibility [31] as well as streaking artifacts and noise amplification.

In COSMOS [32], gradient echo data are acquired under multiple head rotations (r∈[2…N]) with respect to the main magnetic field. This allows the optimization problem to be formulated in matrix form.

$$\begin{pmatrix} d_1(k) \\ \vdots \\ d_N(k) \end{pmatrix} \chi(k) = \begin{pmatrix} \phi_1(k) \\ \vdots \\ \phi_N(k) \end{pmatrix}$$

It also admits the closed form solution $\chi = (\sum_r d_r^T d_r)^{-1} \sum_r d_r^T \phi_r$ by multiplying the above equation by $\sum_r d_r^T(k)$. With increasing number of head orientations, the conditioning of $(\sum_r d_r^T d_r)^{-1}$ improves since the dipole kernel also rotates and diminishing values around the magic angle do not overlap. This enables high image quality but comes at the cost of long scan time as multi-orientation data need to be acquired. This drawback was partly mitigated using fast imaging techniques such as EPI [33]–[35] or Wave-CAIPI [36]; nevertheless, unnatural head positions/orientations remain a challenge for clinical translation.

Over the last decade, several single-orientation QSM reconstructions were proposed where additional regularization $\|R(\vec{\chi})\|_\ell$ is used to improve the image quality. Commonly encountered regularizers utilize $\ell_1$ or $\ell_2$ penalties (such as in MEDI [1] or L2 [37]), which are either applied on the image itself or on its representation using a custom transform. Note that the optimization problem shown below is posed in image space where $D = F^H d F$ applies the forward and inverse Fourier transform F.

$$\min_{\vec{\chi}} \|D\vec{\chi} - \vec{\phi}\|_2^2 + \lambda \|R(\vec{\chi})\|_\ell$$

However, this formulation assumes that the linear susceptibility-to-field relationship is governed by Gaussian noise, whereas the phase noise distribution deviates from this especially in low-SNR regions [38]. This was recognized in nonlinear-MEDI [39], where a nonlinear fidelity term was utilized.

$$\min_{\vec{\chi}} \left\| W \left( e^{iD\vec{\chi}} - e^{i\vec{\phi}} \right) \right\|_2^2 + \lambda \|MG\vec{\chi}\|_1$$

Here the magnitude $W$ serves as a noise-weighting factor as well as allowing the derivation of a binary mask $M$ that weights the gradient $G$. This approach efficiently mitigates artifacts and improves the image appearance. However, the image quality strongly depends on the choice of the regularization parameter $\lambda$ which balances accuracy (data consistency) vs. image smoothness. Another drawback is the time-consuming reconstruction using complex optimizers. This issue was addressed by the recently proposed FANSI algorithm [40] which presents a rapid alternative to nonlinear-MEDI by employing parameter splitting [41], [42]. This provides up to 10-fold computational speed-up but comes at the cost of two additional regularization parameters that need to be tuned manually.

Alternative approaches employ Total Variation (TV) to promote image smoothness and reduce streaking artifacts [43]. However, as TV only takes the first derivative into account, it neglects higher order smoothness and hence assumes that images are piecewise constant, which may lead to over-smoothing and unnatural image appearance. Total Generalized Variation (TGV) lifts this assumption by balancing both first and second derivatives, which is demonstrated to improve the image quality and prevent staircase artifacts [33].

Further improvement in image quality was achieved using single-step reconstruction algorithms [44] which were proposed to mitigate potential error propagation between subsequent procedures along the QSM pipeline. As demonstrated in [44], operators for Laplacian unwrapping and spherical mean value (SMV) background filtering can be directly integrated into the optimization problem. While this increases the computational footprint, it further reduces reconstruction errors when compared to multi-step reconstruction algorithms.

Recent advances in deep learning gained wide-spread attention in the MRI research community. Convolutional neural networks (CNN) were trained to perform the deconvolution based on single-orientation phase data (QSMNet [45]) and provided similar outcome as multi-orientation COSMOS reconstructions. DeepQSM [46] further demonstrated that the mathematical principle of dipole inversion can be learned entirely using synthetic images, and this network generalized to unseen patient data. As such, DeepQSM could potentially circumvent the demand for large amounts of patient training data.

In this contribution, we develop a simple gradient descent optimizer - Nonlinear Dipole Inversion (NDI) - and demonstrate how magnitude weighting and nonlinear formulation act as inherent priors, thus obviating the need for manual parameter tuning. We then expand NDI to learn variational regularizers from training data to further improve the image quality. Ultimately, we leverage Wave-CAIPI encoding to acquire highly accelerated high-resolution data at 7T and evaluate the performance of NDI at 0.5 mm isotropic resolution.

Code/data: https://bit.ly/2RHeiF0

# Method

## Nonlinear Dipole Inversion (NDI)

NDI is based on the nonlinear-MEDI [39] approach, but additional regularization terms are entirely removed and magnitude weighting and nonlinear formulation are exploited as inherent regularizers for the NDI reconstruction.

$$f(\vec{\chi}) = \left\| W \left( e^{iD\vec{\chi}} - e^{i\vec{\phi}} \right) \right\|_2^2$$

This allows an analytical derivation of the gradient $\nabla_{\vec{\chi}} f(\vec{\chi})$ (see Appendix for details) and the application of gradient descent optimization.

$$\nabla_{\vec{\chi}} f(\vec{\chi}) = 2 D^T W^T W \sin(D\vec{\chi} - \vec{\phi})$$

With this, the $t^{th}$ update of the reconstruction becomes

$$\vec{\chi}^{t+1} = \vec{\chi}^t - 2 \sum_{r=1}^{N} D_r^T W_r^T W_r \sin(D_r \vec{\chi}^t - \vec{\phi}_r)$$

where we generalized the formula for multi-orientation reconstruction from $N$-directions, with $W_r$, $\vec{\phi}_r$ and $D_r$ denoting the magnitude, tissue phase and dipole kernel belonging to the $r^{th}$ head rotation. Also, this framework can be easily expanded to allow for Tikhonov regularization by subtracting $2\lambda \vec{\chi}^t$ from the above equation.

## Data acquisition and preparation

We used the QSMNet dataset [45] where 3D-GRE data were acquired on nine subjects using five head orientations with 1 mm isotropic resolution, 256x224x192 matrix, TE/TR=25/33 ms, flip-angle=15°, bandwidth=100 Hz/px and R=2x2 GRAPPA acceleration [47] at 3T.

On a 7T research system (Siemens Healthcare, Erlangen, Germany), we acquired 3D-GRE data at 0.5 mm isotropic resolution on one healthy volunteer using a prototype Wave-CAIPI [36] sequence (3 head orientations, 480x480x360 matrix, TE/TR=19/29 ms, flip angle=25°, bandwidth=100 Hz/px, R=5x3 acceleration, acquisition time TA=5:13 min per orientation). A custom tight-fitted 31-channel head coil [48] (non-product) was used to achieve high-quality imaging; however, this limited the feasible head rotations to shallow angles (0°, 7°, 13°). Additional low-resolution GRE reference scans were acquired for each head orientation to compute coil sensitivity maps using ESPIRiT [49]. The parallel imaging reconstruction was performed offline using MATLAB, where gradient imperfections were corrected in an entirely data-driven fashion using AutoPSF [50] which obviates the need for time-consuming calibration scans.

All multi-orientation data were processed offline using BET brain masking [51], FLIRT registration [52], Laplacian unwrapping [53], and SMV filtering [54]. In order to increase the number of training samples for our deep learning reconstructions, the following data augmentation strategy was performed on the QSMNet dataset: For each of the nine subjects, the multi-orientation tissue phase was first registered to the neutral head orientation, and 5-direction NDI and COSMOS were computed. The resulting ground-truth images were then registered back to each of the four rotated head directions of the original QSMNet dataset. This increased the number of input/ground-truth pairs by a factor of five resulting in a total of 45 datasets from nine subjects.

### Tikhonov regularization in NDI

The NDI optimization may start to fit noise and artifacts to further reduce the cost function if too many iterations are performed. As an alternative to early-stopping, we investigate the effect of Tikhonov regularization. It provides a framework to stabilize the solution of ill-conditioned linear equations and has been successfully applied to related problems such as parallel imaging [55]. We examine the convergence of single- and multi-orientation NDI reconstructions with Tikhonov regularization ($\lambda$=0.0%, $\lambda$=0.1%, $\lambda$=1.0%) by computing the RMSE with respect to 5-direction NDI for different number of iterations throughout the optimization (Fig. 1).

### NDI vs. COSMOS, TKD, L2 and FANSI

Non-regularized NDI and COSMOS were compared for single- and multi-orientation reconstructions (Fig. 2). To prevent over-fitting in NDI, the optimization was stopped after 400 iterations (compare results of Fig. 1). Moreover, single-orientation NDI with Tikhonov regularization ($\lambda$=0.1%) was compared against TKD, L2 and FANSI where the parameters of the latter three were tuned to minimize RMSE with respect to the 5-direction COSMOS data (Fig. 3). In Table 1, RMSE and SSIM are reported with respect to 5-direction COSMOS and NDI. Moreover, the QSM data consistency ($\left\|D\vec{\chi}-\vec{\phi}\right\|_2$) and the computation time in MATLAB are provided for each algorithm.

### Learning a variational regularizer for NDI (VaNDI)

We introduce VaNDI to further improve the reconstruction quality of NDI using a Variational Network (VN) [56] which combines deep learning elements with the nonlinear QSM data model (Fig. 4). This network architecture acts as an unrolled gradient descent algorithm with learned regularizers, where the step sizes, nonlinearities and convolutional filters are estimated during the training phase. We used an L2 loss function to minimize the difference between 1-direction VaNDI and 5-direction NDI reference data, with T=10 iterations, 24 filter kernels (9x9x9), batch size of 1, and 800 epochs. As benchmark of comparison, we used a 3D UNet [57] with the same input/ground truth data as in VaNDI and the following network parameters: depth= 4, 32 filter kernels of size 3x3x3. Both networks were trained on eight volunteers (five head orientations per subject, 40 3D datasets in total), the ninth subject was used for testing. The results of the image quality comparison are summarized in Fig. 5 and Table 2.

### High-resolution NDI at 7T with Wave-CAIPI encoding

We also assessed the performance of NDI on high-resolution GRE data from three head orientations acquired at 7T. To facilitate such high-resolution acquisition in a reasonable timeframe, Wave-CAIPI encoding at R=15-fold acceleration was used. For benchmark of comparison, 3-orientation COSMOS was computed and compared to single- and multi-orientation NDI (see Fig. 6).

## Results

In Figure 1, the effect of early-stopping and Tikhonov regularization is assessed. We computed the RMSE of 1–4-direction NDI with respect to 5-direction NDI for different Tikhonov regularization values. Both for single- and multi-orientation reconstructions, NDI was over-fitting the data when no regularization was used ($\lambda$=0.0%, blue line). While the optimal stopping iteration with the smallest RMSE (blue dashed) varied between the different reconstructions, a fixed small amount of Tikhonov regularization ($\lambda$=0.1%, red line) prevented over-fitting for all observed numbers of directions and subjects (see Appendix).

Figure 2 compares NDI and COSMOS for single- and multi-orientation reconstructions. For small number of head orientations, COSMOS is subject to artifacts as $(\sum_r d_r^T d_r)^{-1}$ is poorly conditioned. NDI could address this and improved the reconstruction quality dramatically even from a single-orientation input. At five head directions, both techniques provided comparable image quality and contrast.

Figure 3 compares single-orientation NDI against parameter-optimized TKD, L2 and FANSI. The regularization in L2 and FANSI leads to blurring and over-smoothing when compared to NDI and TKD, which is also reflected in a larger data consistency error (see Table 1). In contrast, TKD provided sharper images but suffered from more streaking artifacts and contrast reduction, a consequence of the k-space underestimation around the magic angle (see k-space picture). A good trade-off between mitigation of artifacts and image sharpness was achieved by NDI, which led to a medium data consistency error. The RMSE/SSIM metrics computed with respect to 5-direction COSMOS (Table 1) yielded overall comparable results for all algorithms under consideration and did not seem to reflect artifacts and loss of sharpness. Only when compared to 5-dir NDI, the contending algorithms demonstrated inferior RMSE/SSIM.

Figure 5 compares the results of 1-direction VaNDI (deep learning + nonlinear data-fidelity) and UNet. Both approaches overall improved the image quality when compared to single-direction NDI which resulted in better RMSE/SSIM, but increased data consistency error due to the additional regularization. Moreover, slight underestimation of the susceptibility signal in the single direction NDI input data was mitigated by both techniques (e.g. in the basal ganglia). However, while UNet achieved better GM/WM contrast and overall crisper images, it introduced additional artifacts (marked with red arrows), an effect not observed in any of the VaNDI reconstructions.

The performance of NDI and COSMOS was also evaluated at 7T using high-resolution data (0.5 mm isotropic) acquired with the Wave-CAIPI GRE sequence. While 3-direction COSMOS resulted in poor image quality with streaking artifacts (max. head rotations was 13°), 1-direction NDI provided better reconstructions. Further improvement was achieved using 3-direction NDI, where small anatomical features such as blood vessels and U-fibers (zoom-in) were more conspicuous.

## Discussion

We developed a robust and simple dipole inversion technique and demonstrated high-quality reconstructions from an arbitrary number of head orientations.

NDI does not use complicated regularizers (no spatial gradient penalty, TV, etc.) but relies on the inherent regularization effect introduced by magnitude weighting and nonlinear formulation. In addition, either early-stopping or a small amount of Tikhonov regularization can be employed to improve the convergence. A suitable Tikhonov regularization parameter was empirically determined, and robustness among different subjects and number of head orientations was observed (compare Fig. S1). In practice, this enables high-quality NDI reconstructions with small computational footprint and without the need for manual parameter tuning.

In our experimental validation, NDI outperformed COSMOS for small number of head orientations (1–3) where the COSMOS reconstruction is poorly conditioned and numerically unstable. Towards larger number of head orientations, NDI and COSMOS provided comparable image quality and contrast.

Moreover, we demonstrated that NDI matches the RMSE/SSIM metrics from FANSI without the need for parameter tuning, and without the vulnerability of over-smoothing the images. This was mainly observed when RMSE/SSIM was compared to 5-direction COSMOS, as the magnitude weighting imposed a bias in favor of NDI. In terms of model agreement, TKD and NDI achieved better data consistency than L2 and FANSI, as only little regularization was applied which also lead to sharper images. Nevertheless, a caveat of NDI is its longer reconstruction time (~2 min); however, we anticipate further speed-up using more advanced optimization techniques (e.g. non-linear conjugate gradient with backtracking line search [58]). This should result in much faster convergence and reduced reconstruction time.

We also investigated a novel deep learning approach to further refine the image quality by expanding NDI to admit variational regularizers learned from training data. Our VaNDI technique was compared to a UNet architecture, where comparable RMSE/SSIM was observed. However, while UNet achieved slightly better contrast/sharpness, the VaNDI approach was less susceptible to artifacts ("hallucination"), which we believe is a result of the integrated data-fidelity term. While this issue may be mitigated using more training data, VaNDI seems to be a robust alternative in the presence of limited training data availability.

Ultimately, NDI was applied to 7T where GRE data was acquired at 0.5 mm isotropic resolution with Wave-CAIPI encoding (R=15x acceleration) and a custom tight-fitted coil to achieve high-quality imaging. This, however, limited the achievable head rotations to shallow angles which created a difficult dataset for QSM reconstruction, resulting in streaking artifacts in the COSMOS technique. In contrast, the inherent regularization of NDI enabled much better quality even at a single orientation, which further improved with more directions. This – as far as we know – may have revealed iron content in the U-fibers for the first time using in-vivo QSM. In contrast to previous publications [36], we also corrected for imperfections of the Wave gradients in an entirely data-driven fashion (using AutoPSF [50]) without the necessity for time-consuming calibration scans. This enabled high-resolution QSM to become feasible in a much shorter acquisition/reconstruction time and should help pave the way for more frequent usage in the neuroscientific research community.

In conclusion, we developed a simple gradient descent optimizer to perform robust QSM without the need for parameter tuning. We then combined NDI synergistically with deep learning where variational regularizers were learned from training data (VaNDI) to improve the image quality. We further demonstrated feasibility of high-resolution NDI at 7T where Wave-CAIPI was utilized to facilitate highly accelerated acquisitions.

# Appendix

## Analytical gradient derivation for NDI optimization

NDI is based on nonlinear MEDI, where additional regularization terms were removed ($\alpha=0$). This allows an analytical gradient derivation which is provided in this section. NDI aims to minimize the following cost function

$$f(\vec{\chi}) = \left\| W \left( e^{iD\vec{\chi}} - e^{i\vec{\phi}} \right) \right\|_2^2$$

which can be rewritten as

$$= \left( e^{iD\vec{\chi}} - e^{i\vec{\phi}} \right)^H W^T W \left( e^{iD\vec{\chi}} - e^{i\vec{\phi}} \right)$$

Since $D\vec{\chi}$ is constrained to real values, we can use $(D\vec{\chi})^H = (D\vec{\chi})^T$

$$= \left( e^{-i(D\vec{\chi})^T} W^T W e^{iD\vec{\chi}} - e^{-i(D\vec{\chi})^T} W^T W e^{i\vec{\phi}} - e^{-i\vec{\phi}^T} W^T W e^{iD\vec{\phi}} + e^{-i\vec{\phi}^T} W^T W e^{i\vec{\phi}} \right)$$

Rewriting matrix multiplications using index notation yields

$$= \sum_c w_{cc}^2 \left( 2 - e^{i(\sum_b D_{cb}\chi_b - \phi_c)} - e^{-i(\sum_b D_{cb}\chi - \phi_c)} \right)$$

Applying a trigonometric relation further simplifies the term

$$= 2 \sum_c w_{cc}^2 \left( 1 - \cos\left( \sum_b D_{cb}\chi_b - \phi_c \right) \right)$$

Differentiating $f(\chi)$ with respect to $\chi_n$ yields

$$\frac{\partial f}{\partial \chi_n} = 2 \sum_c w_{cc}^2 D_{cn} \cdot \sin\left( \sum_b D_{cb}\chi_b - \phi_c \right)$$

Ultimately, we return to matrix notation and obtain

$$\nabla_{\vec{\chi}} f(\vec{\chi}) = 2 D^T W^T W \sin(D\vec{\chi} - \vec{\phi})$$

# Tables

|  | NDI | TKD | L2 | FANSI |
|---|---|---|---|---|
| RMSE w.r.t. 5-dir NDI | 0.567 | 0.680 | 0.710 | 0.639 |
| RMSE w.r.t. 5-dir COSMOS | 0.599 | 0.607 | 0.626 | 0.568 |
| SSIM w.r.t. 5-dir NDI | 0.948 | 0.921 | 0.900 | 0.916 |
| SSIM w.r.t. 5-dir COSMOS | 0.940 | 0.942 | 0.925 | 0.941 |
| Data consistency $\|D\vec{\chi} - \vec{\phi}\|_2$ | 0.505 | 0.464 | 0.562 | 0.563 |
| Reconstruction time [s] | 133 | 0.12 | 0.14 | 17 |

**Table 1: RMSE/SSIM was computed w.r.t. 5-direction NDI and COSMOS. Moreover, the data consistency and reconstruction times are reported.**

|  | NDI | VaNDI | UNet |
|---|---|---|---|
| RMSE w.r.t. 5-dir NDI | 0.712 | 0.575 | 0.578 |
| SSIM w.r.t. 5-dir NDI | 0.845 | 0.927 | 0.910 |
| Data consistency $\|D\vec{\chi} - \vec{\phi}\|_2$ | 0.533 | 0.663 | 0.673 |

**Table 2: RMSE/SSIM (w.r.t. 5-direction NDI) and data consistency are provided for single-direction NDI, VaNDI and UNet.**

# Figures

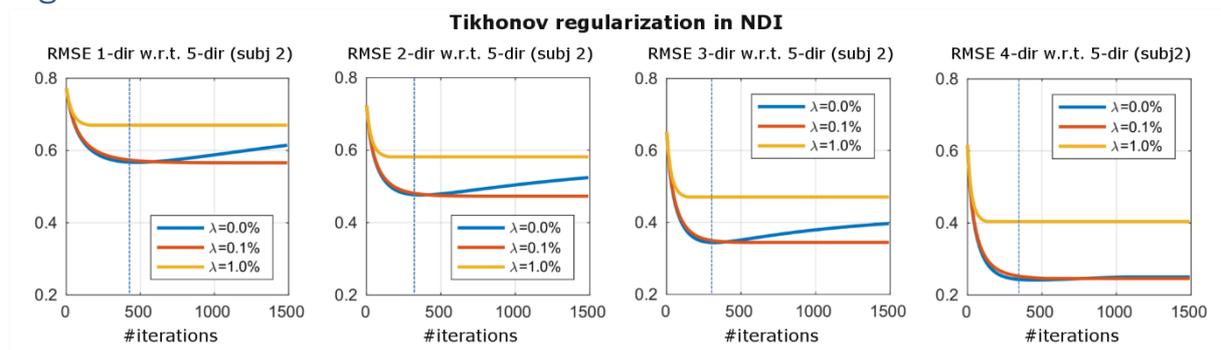

Figure 1: NDI over-fits the data ($\lambda$=0.0%, blue line), if not stopped early (blue dashed line). A small amount of Tikhonov regularization ($\lambda$=0.1%, red line) mitigated this issue robustly for all observed number of directions and subjects (see Appendix).

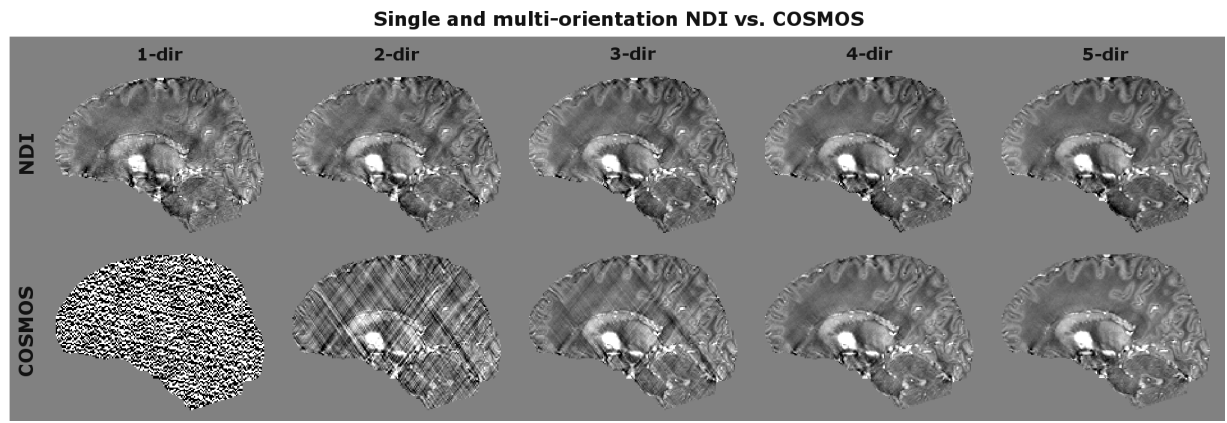

Figure 2: Comparison of NDI vs. COSMOS for various numbers of head directions. NDI significantly reduced streaking artifacts and provided good results even at a single head orientation.

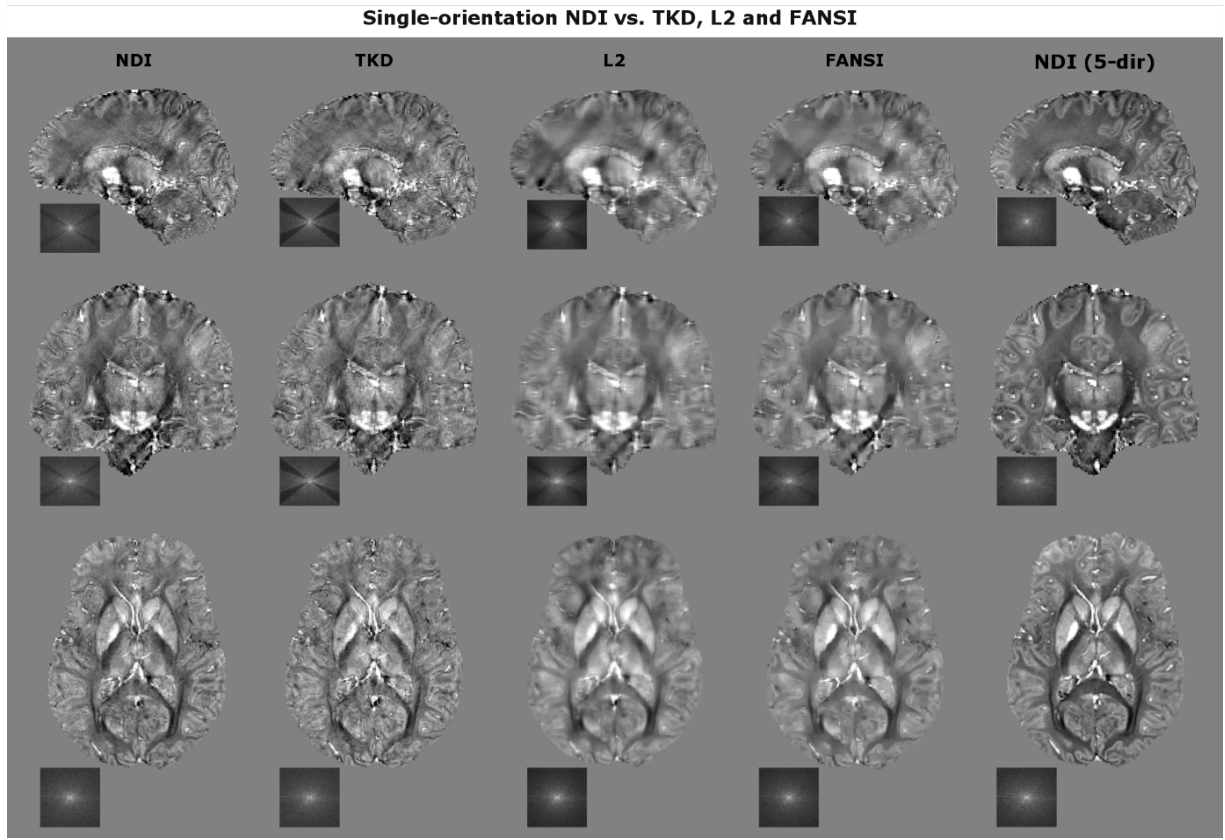

Figure 3: Comparison of NDI vs. optimized TKD, L2 and FANSI. NDI mitigated streaking artifacts (observed in TKD) while preventing large blurring and over-smoothing as observed in L2 and FANSI.

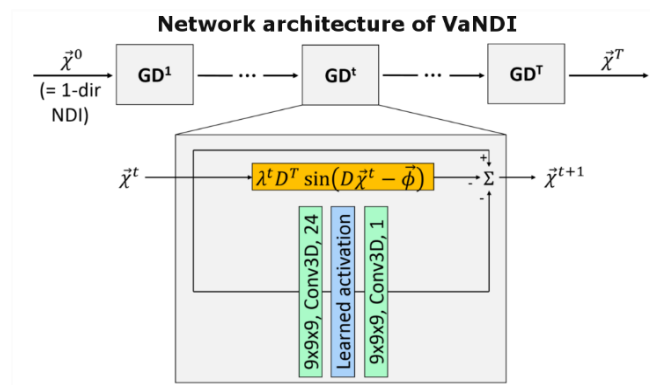

Figure 4: VaNDI acts as an unrolled gradient descent algorithm combining deep learning and nonlinear data-fidelity. In the CNN part of the network, 3D convolutions (green) and nonlinear activation (blue) are learned for each gradient descent step $GD^t$. Moreover, a data-fidelity term (orange) is integrated into the network to ensure agreement with the QSM model.

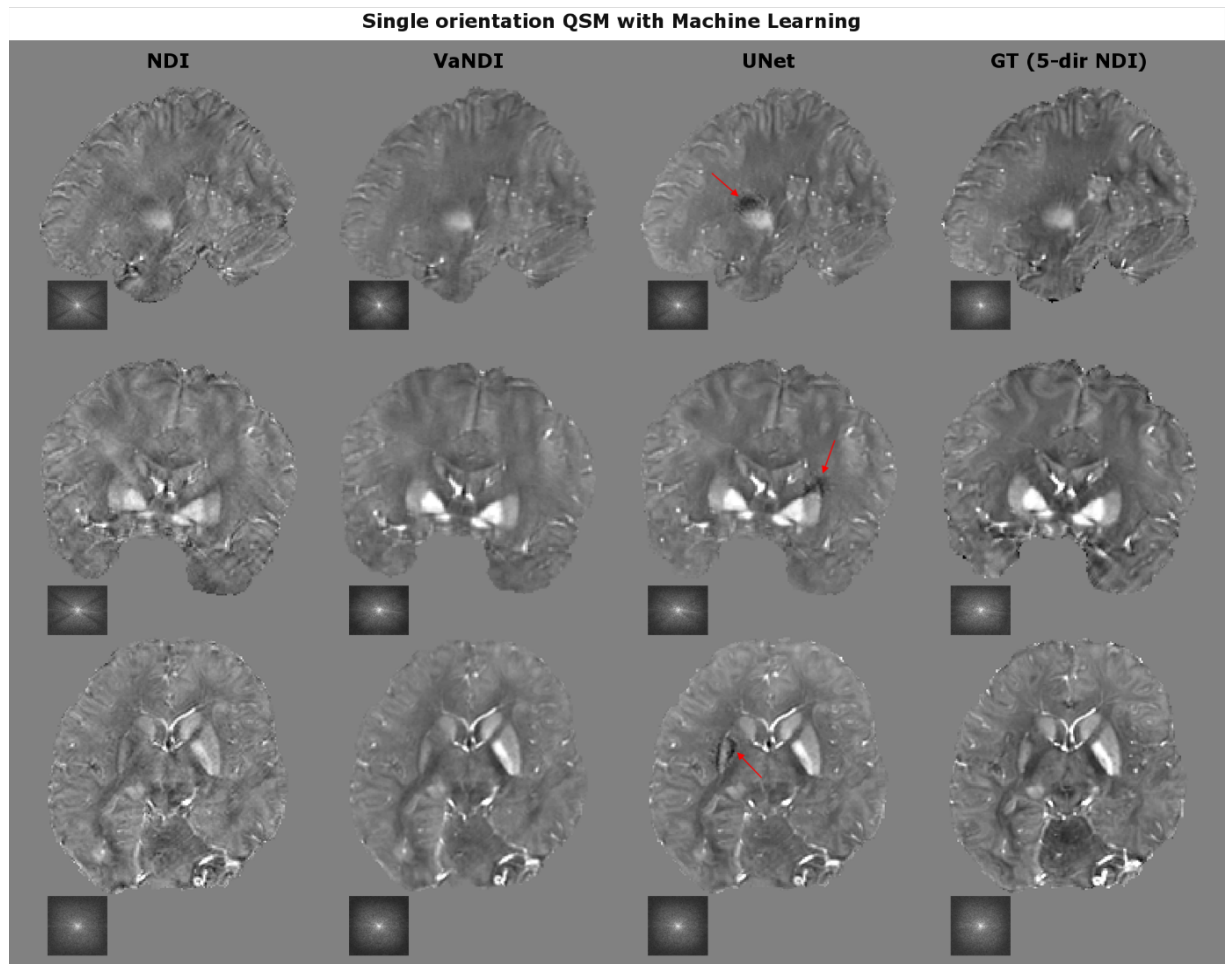

Figure 5: VaNDI and UNet further improved the image quality of single-direction NDI by reducing artifacts and improving the image contrast. However, while UNet better preserved the image sharpness, it introduced additional artifacts (red arrow) which was not observed in any of the VaNDI reconstructions.

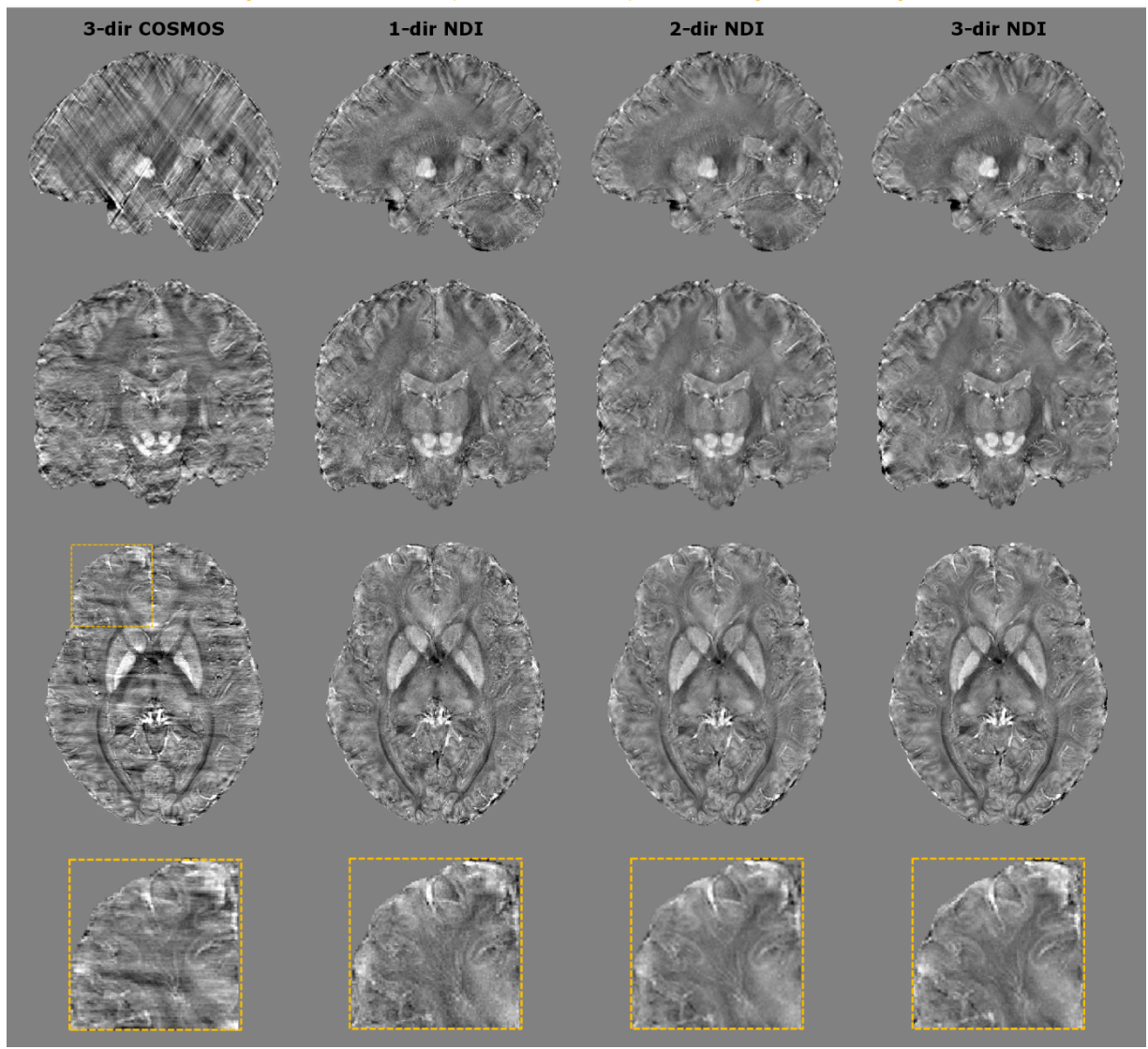

Figure 6: High-resolution QSM data was generated from multi-orientation GRE scans at 0.5 mm isotropic resolution and R=15x acceleration using Wave-CAIPI encoding. 3-direction COSMOS resulted in streaking artifacts, which was much improved using NDI even at a single orientation. The zoom-ins reveal fine anatomical features such as the U-fibers which are best seen in the 3-direction NDI reconstruction.

# Supplementary Material

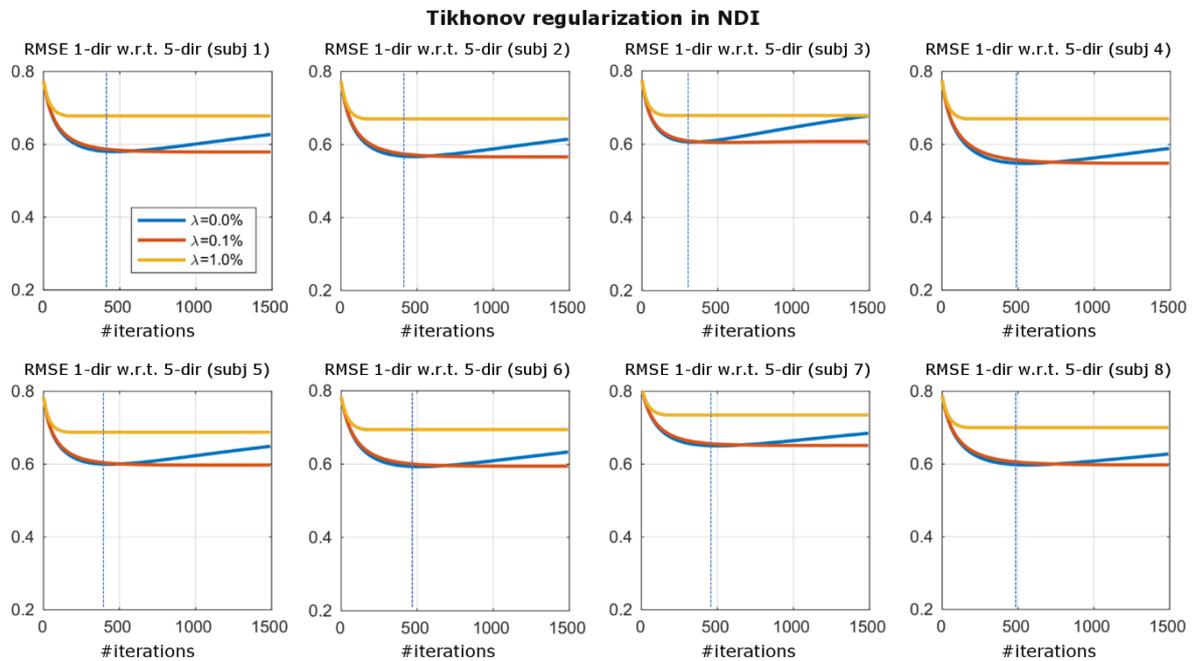

S1: NDI may lead to overfitting ($\lambda = 0$, blue line), if not stopped early (blue dashed line), which can be mitigated using a small amount of Tikhonov regularization. As demonstrated, $\lambda$=0.1% robustly stabilized the reconstruction for all observed subjects.